\documentclass[iop]{emulateapj}

\usepackage{hyperref}  
\usepackage{breakurl}  

\newcommand\kms{\ifmmode{\rm km\thinspace s^{-1}}\else km\thinspace s$^{-1}$\fi}
\newcommand\btvul{BT~Vul}

\shortauthors{Torres et al.}
\shorttitle{\btvul}

\begin{document}
\submitted{Accepted for publication in The Astrophysical Journal}

\title{Absolute dimensions of the unevolved F-type eclipsing binary BT
  Vulpeculae}
\shorttitle{The eclipsing binary \btvul}

\author{
Guillermo Torres\altaffilmark{1},
Claud H.\ Sandberg Lacy\altaffilmark{2},
Francis C. Fekel\altaffilmark{3}, and
Matthew W. Muterspaugh\altaffilmark{4,5}
}

\altaffiltext{1}{Center for Astrophysics \textbar\ Harvard \&
  Smithsonian, 60 Garden Street, Cambridge, MA 02138, USA;
  gtorres@cfa.harvard.edu}

\altaffiltext{2}{Physics Department, University of Arkansas,
  Fayetteville, AR 72701, USA}

\altaffiltext{3}{Center of Excellence in Information Systems,
  Tennessee State University, Nashville, TN 37209, USA}

\altaffiltext{4}{Fairborn Observatory, 1327 Duquesne Rd, Patagonia, AZ
  85624}

\altaffiltext{5}{Current address: Division of Science, Technology, and
  Mathematics, Columbia State Community College, 1665 Hampshire Pike,
  Columbia, TN 38401}

\begin{abstract}

We report extensive differential $V$-band photometry and
high-resolution spectroscopy for the 1.14~day, detached, double-lined
eclipsing binary \btvul\ (F0+F7). Our radial-velocity monitoring and
light curve analysis lead to absolute masses and radii of $M_1 =
1.5439 \pm 0.0098~\mathcal{M}_{\sun}^{\rm N}$ and $R_1 = 1.536 \pm
0.018~\mathcal{R}_{\sun}^{\rm N}$ for the primary, and $M_2 = 1.2196
\pm 0.0080~\mathcal{M}_{\sun}^{\rm N}$ and $R_2 = 1.151 \pm
0.029~\mathcal{R}_{\sun}^{\rm N}$ for the secondary. The effective
temperatures are $7270 \pm 150$~K and $6260 \pm 180$~K,
respectively. Both stars are rapid rotators, and the orbit is
circular. A comparison with stellar evolution models from the MIST
series shows excellent agreement with these determinations, for a
composition of ${\rm [Fe/H]} = +0.08$ and an age of 350~Myr. The two
components of \btvul\ are very near the zero-age main sequence.

\end{abstract}

\section{Introduction}
\label{sec:introduction}

The discovery of the photometric variability of \btvul\ (HD~340072,
TYC~2164-161-1, Gaia DR2 1836032243422750464; $V = 11.49$, SpT F0+F7)
was made by S.\ Beljawsky \citep[see][]{Guthnick:1939}, although the
original period given (3.988~days) was incorrect. Other than
measurements of the times of eclipse since its discovery, no detailed
studies of this short-period system ($P = 1.14$~days) have appeared in
the literature. In this paper we report the first systematic
photometric and spectroscopic monitoring of the binary, leading to a
full determination of its physical properties.

Section~\ref{sec:observations} presents our observations, beginning in
Section~\ref{sec:timings} with a determination of a highly precise
ephemeris based on more than six decades of timing observations.
Section~\ref{sec:spectroscopy} then reports our spectroscopic
observations of \btvul, with the derivation of the radial velocities
and spectroscopic orbital elements. Our extensive photometric $V$-band
measurements are described in Section~\ref{sec:photometry}. The
analysis of the lightcurves is found in Section~\ref{sec:analysis},
and is followed in Section~\ref{sec:dimensions} by a determination of
the absolute dimensions of the system, and in Section~\ref{sec:models}
with a comparison of the physical properties with current models of
stellar evolution. Final remarks appear in Section~\ref{sec:remarks}.

\section{Observations}
\label{sec:observations}

\subsection{Eclipse timings}
\label{sec:timings}

Times of minimum light have been gathered for \btvul\ since 1953 using
photographic, visual, and photoelectric/CCD techniques. The available
measurements are collected in Table~\ref{tab:minima}, which contains
115 timings for the primary minimum and 40 for the secondary over more
than 65 years. Many of them are based on the differential photometric
observations we describe in this paper.

A linear ephemeris was derived by weighted least squares, in which the
published uncertainties, when available, were scaled in order to
achieve reduced $\chi^2$ values near unity for each type of eclipse
and each type of observation. For measurements with no reported
uncertainties we adopted suitable values to the same end (see
Table~\ref{tab:minima}).  The linear ephemeris obtained is:
$${\rm Min~I (HJD)} = 2,\!453,\!479.931997 (92) + 1.141200674 (50) E,$$
with the uncertainties indicated in parentheses. Independent periods
determined from the primary and secondary minima do not differ
significantly. A solution allowing for separate reference times of
primary and secondary minima resulted in a phase difference of
$0.49987 \pm 0.00014$, suggesting negligible eccentricity.  For the
rest of this work we have therefore assumed the orbit to be circular.

\setlength{\tabcolsep}{2pt}  
\begin{deluxetable}{lcccccc}
\tablewidth{0pc}
\tablecaption{Times of Minimum Light for \btvul\ \label{tab:minima}}
\tablehead{
\colhead{HJD} &
\colhead{} &
\colhead{} &
\colhead{$\sigma$} &
\colhead{$O-C$} &
\colhead{} &
\colhead{}
\\
\colhead{(2,400,000+)} &
\colhead{Year} &
\colhead{Epoch} &
\colhead{(days)} &
\colhead{(days)} &
\colhead{Ecl} &
\colhead{Type}
}
\startdata
 34740.28    &  1953.9912  &  $-$16421.0  &  0.0063  &  +0.004269  &  1  &   PG \\
 35402.18    &  1955.8034  &  $-$15841.0  &  0.0063  &  +0.007879  &  1  &   PG \\
 43351.786   &  1977.5682  &   $-$8875.0  &  0.0100  &  +0.009984  &  1  &   V  \\
 43779.735   &  1978.7399  &   $-$8500.0  &  0.0100  &  +0.008731  &  1  &   V  \\
 44022.812   &  1979.4054  &   $-$8287.0  &  0.0100  &  +0.009988  &  1  &   V 
\enddata
\tablecomments{`Ecl' is 1 for the primary and 2 for the secondary
  minimum; `Type' is PG, V, and PE for the photographic, visual, and
  photoelectric/CCD techniques. The `Epoch' is counted from the
  reference time of primary minimum in this section. For measurements
  with published uncertainties the errors reported here have been
  multiplied by scale factors of 3.35 and 2.75 for the photoelectric
  timings of the primary and secondary minima (see text), and scale
  factors of 4.35 and 1.28 for the visual observations. Measurements
  with no published errors have been assigned values of 0.0063 days,
  0.010 days, and 0.0058 days for the photographic, visual, and
  photoelectric techniques, respectively. Sources for the individual
  measurements are listed at
  \url{http://var2.astro.cz/EN/brno/eclipsing_binaries.php} and
  \url{https://www.bav-astro.eu/index.php/veroeffentlichungen/service-for-scientists/lkdb-engl}.
  (This table is available in its entirety in machine-readable form.)}
\end{deluxetable}
\setlength{\tabcolsep}{6pt}  

\subsection{Spectroscopy}
\label{sec:spectroscopy}

Spectroscopic monitoring of \btvul\ was carried out at two facilities.
Observations at the Center for Astrophysics (CfA) began in June of
2010 and continued until October of 2015. They were made with the
Tillinghast Reflector Echelle Spectrograph
\citep[TRES;][]{Szentgyorgyi:2007, Furesz:2008}, a fiber-fed,
bench-mounted instrument attached to the 1.5m Tillinghast reflector at
the Fred L.\ Whipple Observatory on Mount Hopkins (Arizona, USA). A
total of 39 spectra were gathered at a resolving power of $R \approx
44,\!000$, and cover the wavelength region 3800--9100~\AA\ in 51
orders. For the order centered at $\sim$5187~\AA, which contains the
\ion{Mg}{1}~b triplet, the signal-to-noise ratios range from 24 to 65
per resolution element of 6.8~\kms. The wavelength reference was
provided by exposures of a thorium-argon lamp before and after each
science frame, and the reductions were carried out with a dedicated
pipeline.

\setlength{\tabcolsep}{3pt}  
\begin{deluxetable}{crrccc}
\tablewidth{0pc}
\tablecaption{CfA Radial Velocities for \btvul\ \label{tab:rvs_cfa}}
\tablehead{
\colhead{HJD} &
\colhead{$RV_1$} &
\colhead{$RV_2$} &
\colhead{$\sigma_1$} &
\colhead{$\sigma_2$} &
\colhead{Orbital}
\\
\colhead{(2,400,000+)} &
\colhead{(\kms)} &
\colhead{(\kms)} &
\colhead{(\kms)} &
\colhead{(\kms)} &
\colhead{Phase}
}
\startdata
 55368.9509  &   $-$139.96  &   136.65    &   4.56  &  4.60  &  0.29074 \\
 55527.5607  &   $-$142.87  &   135.21    &   3.54  &  3.57  &  0.27576 \\
 55850.5866  &   $-$129.49  &   121.77    &   2.50  &  2.52  &  0.33369 \\
 55903.5554  &      103.22  &  $-$177.41  &   3.99  &  4.02  &  0.74866 \\
 56024.0077  &   $-$143.75  &   133.62    &   1.86  &  1.87  &  0.29741 
\enddata
\tablecomments{Orbital phases are counted from the reference time of
  primary eclipse. (This table is available in its entirety in
  machine-readable form.)}
\end{deluxetable}
\setlength{\tabcolsep}{6pt}  

For the radial velocity measurements we used the two-dimensional
cross-correlation algorithm TODCOR \citep{Zucker:1994}, with separate
templates for the primary and secondary selected from a large library
of pre-computed synthetic spectra based on model atmospheres by
R.\ L.\ Kurucz, and a line list tuned to better match the spectra of
real stars \citep[see][]{Nordstrom:1994, Latham:2002}. The synthetic
spectra cover a limited wavelength region of about 300~\AA\ centered
around 5187~\AA, and the velocity measurements were made using the
central 100~\AA, which contains most of the information on the
velocities. The main template parameters (effective temperature,
$T_{\rm eff}$, and rotational broadening, $v \sin i$) were determined
by running extensive grids of cross-correlations as described by
\cite{Torres:2002}, with adopted surface gravities of $\log g = 4.0$
and 4.5 for the primary and secondary, which are the closest in our
grid to the final values derived later. Solar metallicity was assumed
for these measurements. For the primary we obtained $T_{\rm eff} =
7100$~K and $v \sin i = 70~\kms$. Estimated uncertainties are 100~K
and 3~\kms, respectively, based on the scatter from the individual
spectra conservatively increased to account for possible systematic
errors. For the secondary we obtained $v \sin i = 56 \pm 4~\kms$.
However, the significant rotational broadening of both stars together
with the faintness of the secondary (see below) conspire to make it
difficult to establish a reliable temperature for that star. For this
we therefore made use of results from the light curve analysis
described later. Specifically, we used the surface brightness ratio,
which provides a good measure of the relative temperature. As
described in more detail in Section~\ref{sec:dimensions}, this then
leads to a secondary temperature of 6260~K.  The primary/secondary
template parameters in our grid closest to the above values are
7000\thinspace /\thinspace 6250~K for the temperatures, and
70\thinspace /\thinspace 55~\kms\ for the rotational broadening.

Previous experience at CfA has shown that the raw radial velocities
can sometimes be affected by systematic errors that may arise due to
lines shifting in and out of the spectral window as a function of
orbital phase, or due to the barycentric velocity of the Earth
\citep[see][]{Latham:1996, Torres:1997}.  Following these authors we
applied corrections for this effect based on simulations, the details
of which may be found in those papers.  In the case of the primary the
adjustments are very small ($< 0.4~\kms$), but they can reach
2~\kms\ for the secondary.  The resulting radial velocities in the
heliocentric frame, including corrections, are listed in
Table~\ref{tab:rvs_cfa} along with their uncertainties.  The flux
ratio we measure using TODCOR is $\ell_2/\ell_1 = 0.28 \pm 0.02$, at
the mean wavelength of our observations (5187~\AA).

Further spectroscopic observations of \btvul\ were obtained from 2011
November through 2018 September at Fairborn Observatory, which is
situated in southeast Arizona near Washington Camp. The observations
were acquired with the Tennessee State University 2m Astronomical
Spectroscopic Telescope (AST) and a fiber-fed echelle spectrograph
\citep{Eaton:2007}. The detector was a Fairchild 486 CCD consisting of
a 4K~$\times$~4K array of 15\thinspace $\micron$ pixels.  This
spectrograph and CCD combination resulted in 48 orders covering a
wavelength range of 3800--8260~\AA\ \citep{Fekel:2013}.  The faintness
of the star required our largest diameter fiber that produced a
spectral resolution of 0.4~\AA. As a result, our 52 AST spectra have a
resolving power of 15,000 at 6000~\AA, and an average signal-to-noise
ratio of 35.

\setlength{\tabcolsep}{4pt}  
\begin{deluxetable}{crrc}
\tablewidth{0pc}
\tablecaption{Fairborn Radial Velocities for \btvul\ \label{tab:rvs_fairborn}}
\tablehead{
\colhead{HJD} &
\colhead{$RV_1$} &
\colhead{$RV_2$} &
\colhead{Orbital}
\\
\colhead{(2,400,000+)} &
\colhead{(\kms)} &
\colhead{(\kms)} &
\colhead{Phase}
}
\startdata
 55867.7441  &   $-$115.7  &     97.0  &    0.36829 \\
 56010.9947  &       53.1  & $-$122.1  &    0.89451 \\
 56023.9539  &   $-$151.9  &    136.7  &    0.25027 \\
 56058.9383  &       53.5  & $-$107.2  &    0.90605 \\
 56060.8952  &       67.2  & $-$129.2  &    0.62082 
\enddata
\tablecomments{Typical uncertainties for the velocities are 3.0~\kms\ for
the primary and 3.4~\kms\ for the secondary. Orbital phases are
counted from the reference time of primary eclipse. (This table is
available in its entirety in machine-readable form.)}
\end{deluxetable}
\setlength{\tabcolsep}{6pt}  

A general description of the reduction of the AST spectra was provided
by \cite{Fekel:2009}. In particular, for \btvul\ a solar line list of
168 mostly neutral Fe lines was used in the wavelength region
4920--7100~\AA. The individual lines were fitted with a rotational
broadening function \citep{Lacy:2011}. A value of 0.6~\kms, determined
from our unpublished measurements of IAU velocity standards, has been
added to the AST velocities to bring them into accord with the results
of \cite{Scarfe:2010}. The corrected AST velocities are listed in
Table~\ref{tab:rvs_fairborn}, and have typical uncertainties of
3.0~\kms\ for the primary and 3.4~\kms\ for the secondary.

Rotational broadening fits in the 23 best AST spectra resulted in $v
\sin i$ values of $70 \pm 3~\kms$ and $56 \pm 3~\kms$ for the primary
and secondary stars, respectively.  Similarly, the average
secondary-to-primary equivalent width ratio was determined to be $0.58
\pm 0.01$.  To convert this value to a true flux ratio, we applied a
correction for the difference in line blocking determined by measuring
equivalent widths for some 150 lines in a synthetic spectrum for the
primary star and another for the secondary. These synthetic spectra
were were calculated with the {\tt SPECTRUM\/} code
\citep{Gray:1994}\footnote{\url{http://www.appstate.edu/~grayro/spectrum/spectrum.html}},
and were both normalized to a continuum of unity. With the resulting
correction factor of $1.82 \pm 0.10$, the final spectroscopic flux
ratio for \btvul\ at 6000~\AA\ is $\ell_2/\ell_1 = 0.32 \pm 0.02$.

\begin{figure}
\epsscale{1.15}

\plotone{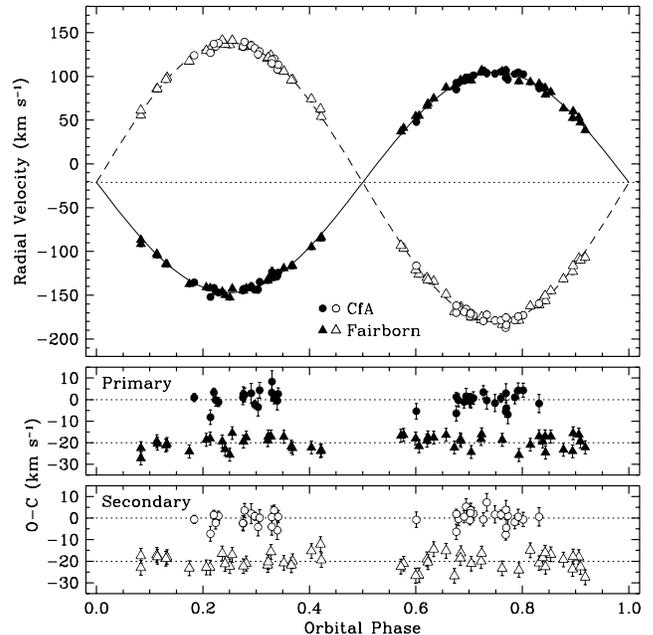}

\figcaption{CfA and Fairborn radial-velocity measurements for
  \btvul\ along with our best fit model that includes times of minimum
  light. Error bars are too small to be visible on the scale of the
  top panel. The dotted line marks the center-of-mass velocity of the
  system.  Residuals and error bars are shown at the bottom, with the
  ones from Fairborn shifted downward by 20~\kms\ for
  clarity. \label{fig:spectroscopy}}

\end{figure}

Independent spectroscopic orbital solutions using the CfA and Fairborn
measurements give the results presented in Table~\ref{tab:orbits}, in
which the period has been held fixed at the value from
Section~\ref{sec:timings}. Slight differences are seen in the velocity
semiamplitudes of both stars, with the CfA values being smaller,
although the values are consistent within 1.3 times their combined
uncertainties.  The CfA solution indicates a small systematic offset
between the primary and secondary velocities ($\Delta RV = -1.13 \pm
0.74$, in the sense primary minus secondary), which can sometimes
occur because of a mismatch between the adopted cross-correlation
templates for TODCOR and the spectra of the real stars. Experiments
with a range of different templates did not reduce the effect. In this
case, however, the shift is not large enough to affect the
semiamplitudes significantly. The Fairborn velocities show no such
offset. For the final solution we used the two data sets together,
solving also for a systematic zero-point difference between Fairborn
and CfA. The results are listed in the last column of the table, and
the observations along with the model may be seen in
Figure~\ref{fig:spectroscopy}. A joint solution allowing for
  eccentricity resulted in a value consistent with zero ($e = 0.0004
  \pm 0.0023$), supporting our assumption of a circular orbit.

\setlength{\tabcolsep}{6pt}  
\begin{deluxetable*}{lccc}
\tablewidth{0pc}
\tablecaption{Spectroscopic Orbital Solutions for \btvul\ \label{tab:orbits}}
\tablehead{
\colhead{~~~~~~~~~~~~~~~~~~~Parameter~~~~~~~~~~~~~~~~~~~} &
\colhead{CfA} &
\colhead{Fairborn} &
\colhead{Combined}
}
\startdata
$K_1$ (\kms)\dotfill                          &   125.46 (52)    &    126.45 (52)     &      125.92 (37)    \\
$K_2$ (\kms)\dotfill                          &   159.32 (52)    &    159.60 (61)     &      159.42 (39)    \\
$\gamma$ (\kms)\dotfill                       &  $-20.96$ (50)   &   $-20.45$ (42)    &     $-21.08$ (48)   \\
Min~I (${\rm HJD} - 2,400,000$)\dotfill       & 53479.9332 (14)  &  53479.93152 (67)  &   53479.931995 (66) \\
CfA prim/sec $\Delta RV$ (\kms)\dotfill       &  $-1.13$ (74)    &     \nodata        &     $-1.38$ (68)    \\
Fairborn prim/sec $\Delta RV$ (\kms)\dotfill  &    \nodata       &   $-0.06$ (64)     &     $-0.02$ (64)    \\
CfA $-$ Fairborn $\Delta RV$ (\kms)\dotfill   &    \nodata       &     \nodata        &     $-0.66$ (64)    \\ \\ [-1.5ex]
\noalign{\hrule} \\ [-1ex]
\multicolumn{4}{c}{Derived quantities} \\ [1ex]
\noalign{\hrule} \\ [-1.5ex]
$q \equiv M_2/M_1$\dotfill                    &   0.7875 (41)    &    0.7923 (45)     &      0.7899 (30)    \\
$M_1 \sin^3 i$ ($M_{\sun}$)\dotfill           &   1.528 (12)     &    1.544 (14)      &      1.5348 (90)    \\
$M_2 \sin^3 i$ ($M_{\sun}$)\dotfill           &   1.203 (10)     &    1.223 (11)      &      1.2124 (74)    \\
$a \sin i$ ($R_{\sun}$)\dotfill               &   6.424 (17)     &    6.453 (18)      &       6.437 (12)    \\ \\ [-1.5ex]
\noalign{\hrule} \\ [-1ex]
\multicolumn{4}{c}{Other quantities pertaining to the fits} \\ [1ex]
\noalign{\hrule} \\ [-1.5ex]
$\sigma_1$ (\kms)\dotfill                     &     3.14         &       3.00         &      3.00 , 3.02    \\
$\sigma_2$ (\kms)\dotfill                     &     3.16         &       3.52         &      3.02 , 3.45    \\
$N_{\rm obs}$\dotfill                         &      39          &        52          &        39 + 52      \\
Time span (days)\dotfill                      &     1929         &       2509         &         3008        
\enddata

\tablecomments{The orbital period has been held fixed according to the
  ephemeris of Section~\ref{sec:timings}. Uncertainties are indicated
  in parentheses in units of the last significant place.}

\end{deluxetable*}
\setlength{\tabcolsep}{6pt}

\subsection{Photometry}
\label{sec:photometry}

Two web-based telescopes were used to gather differential photometric
observations of \btvul: the URSA WebScope \citep[see][]{Torres:2009}
and the NFO WebScope \citep[see][]{Grauer:2008}. The measurement
apertures were $13 \times 13$ pixels square (30 arcsec square) for the
URSA images, and $27 \times 27$ pixels square (22 arcsec square) for
the NFO images. We measured a total of 5451 URSA images and 6018 NFO
images, all taken through a Bessel $V$ filter.  \btvul\ (`var') and
three comparison stars (TYC 2164-0403-1, TYC 2164-0974-1, and TYC
2164-0904-1) were measured in each image. The three comparison stars
appeared to be constant in brightness at the level of 0.010~mag for
the URSA images and 0.007~mag for the NFO images, so the measurement
$\Delta V$ that was used in this analysis was `var$-$comps', where
`comps' is the sum of the brightness of all three comparison
stars. A total of 102 nights of data were gathered for URSA, and 197
nights for NFO.  The measurements are listed in Tables~\ref{tab:ursa}
and \ref{tab:nfo}.

\setlength{\tabcolsep}{6pt}  
\begin{deluxetable}{cc}
\tablewidth{0pc}
\tablecaption{URSA Observations of \\ \btvul.\label{tab:ursa}}
\tablehead{
\colhead{HJD} & \colhead{$\Delta V$}
\\
\colhead{(2,400,000+)} & \colhead{(mag)}
}
\startdata
  52831.67233 &  1.656 \\
  52831.67423 &  1.682 \\
  52831.67611 &  1.687 \\
  52831.67799 &  1.716 \\
  52831.67988 &  1.750 
\enddata
\tablecomments{(This table is available in its entirety in machine-readable form.)}
\end{deluxetable}
\begin{deluxetable}{cc}
\tablewidth{0pc}
\tablecaption{NFO Observations of \\ \btvul.\label{tab:nfo}}
\tablehead{
\colhead{HJD} & \colhead{$\Delta V$}
\\
\colhead{(2,400,000+)} & \colhead{(mag)}
}
\startdata
  53479.88217 &  1.712 \\
  53479.88499 &  1.753 \\
  53479.88777 &  1.783 \\
  53479.89060 &  1.800 \\
  53479.89337 &  1.834 
\enddata
\tablecomments{(This table is available in its entirety in machine-readable form.)}
\end{deluxetable}
\setlength{\tabcolsep}{6pt}  

As we have noted in the past \citep[see, e.g.,][]{Lacy:2008,
  Torres:2014}, these telescopes suffer from systematic shifts of a
few hundredths of a magnitude in the photometric zero point from night
to night, which remain after the reductions. While the effect is
noticeable for the NFO WebScope, it is much less so for URSA. This
adds scatter to the light curves that is addressed below.

\section{Light curve analysis}
\label{sec:analysis}

For the solution of the light curves we used the {\tt eb} code of
\cite{Irwin:2011}, based on the Nelson-Davis-Etzel binary model
\citep{Etzel:1981, Popper:1981}. This model approximates the star
shapes as biaxial spheroids for calculating proximity effects, and is
adequate for well-detached systems in which the stars are nearly
spherical, as is the case here (see below). We solved for the orbital
period ($P$) and reference epoch of primary eclipse ($T_0$), the sum
of the relative radii normalized by the semimajor axis ($r_1+r_2$),
the radius ratio ($k \equiv r_2/r_1$), the central surface brightness
ratio ($J \equiv J_2/J_1$), the cosine of the inclination angle ($\cos
i$), and the magnitude level at first quadrature ($m_0$).  For the
limb darkening prescription we adopted the linear law, with a
coefficient $u$ for each star. Experiments with a quadratic law gave
no improvement. The gravity darkening coefficients were adopted from
\cite{Claret:2011}, and are $y_1 = 0.148$ and $y_2 = 0.327$ for the
primary and secondary, interpolated to our final temperatures, surface
gravities, and metallicity reported below.  The bolometric reflection
albedos ($A_1$, $A_2$) were set to 0.5, and the mass ratio was fixed
to our spectroscopic value, $q = 0.7899$.

A cone search of the {\it Gaia}/DR2 catalog \citep{Gaia:2018} around
the position of \btvul\ revealed about a dozen nearby stars within the
photometric apertures used for URSA and NFO, but none are bright
enough to contaminate the photometry in a significant way.
Nevertheless, as a precaution we included third light ($\ell_3$) as an
additional parameter in our fits. We did this both to account for the
possible presence of closer companions than {\it Gaia\/} can resolve,
and also because of the likelihood of a third, physical component in
the \btvul\ system, given that the vast majority of short-period
binaries appear to have them \citep[see][]{Tokovinin:2006}. Third
light is defined here such that $\ell_1 + \ell_2 + \ell_3 = 1$, and
the values for the primary and secondary for this normalization
correspond to the light at first quadrature.

Our fits were carried out within a Markov chain Monte Carlo framework
using the {\tt emcee} code of
\cite{Foreman-Mackey:2013}.\footnote{\url{https://github.com/dfm/emcee}}
We used 100 walkers with chain lengths of 10,000 each, discarding
  5,000 links as burnin (for a total remaining $5 \times 10^5$
  samples). Priors were uniform for most parameters, with suitable
broad ranges in each case.  Convergence was checked both by visual
examination of the chains and by requiring a Gelman-Rubin statistic of
1.05 or smaller for all parameters \citep{Gelman:1992}. As a final
adjustable parameter we included a multiplicative scale factor $f$ for
the photometric errors, which we solved for self-consistently and
simultaneously with the other orbital quantities
\citep[see][]{Gregory:2005}. We allowed $f$ to be different for URSA
and NFO, as the latter photometry shows more scatter. The prior for
$f$ was assumed to be log-uniform. The initial error assumed for the
photometric measurements was 0.017~mag, estimated from the
out-of-eclipse scatter in the URSA light curve. Formal 68.3\%
  confidence intervals for all parameters were derived directly from
  the posterior distributions.

We performed independent analyses using the URSA and NFO photometry,
the results of which are presented in Table~\ref{tab:mcmc}.
Preliminary solutions produced limb-darkening coefficients near unity
for the primary and near zero for the secondary, which are very
different from their theoretically predicted values and seem
unrealistic. We suspect this may be caused by the systematic offsets
in the photometry from night to night mentioned in
Section~\ref{sec:photometry}, which represent non-Gaussian errors that
can overwhelm the rather subtle effect limb darkening has on the shape
of the light curves. For the remainder of the analysis we therefore
chose to hold the limb-darkening coefficients fixed following the
tabulation of \cite{Claret:2011}, at values of $u_1 = 0.548$ and $u_2
= 0.609$.

\begin{figure}
\epsscale{1.15}
\plotone{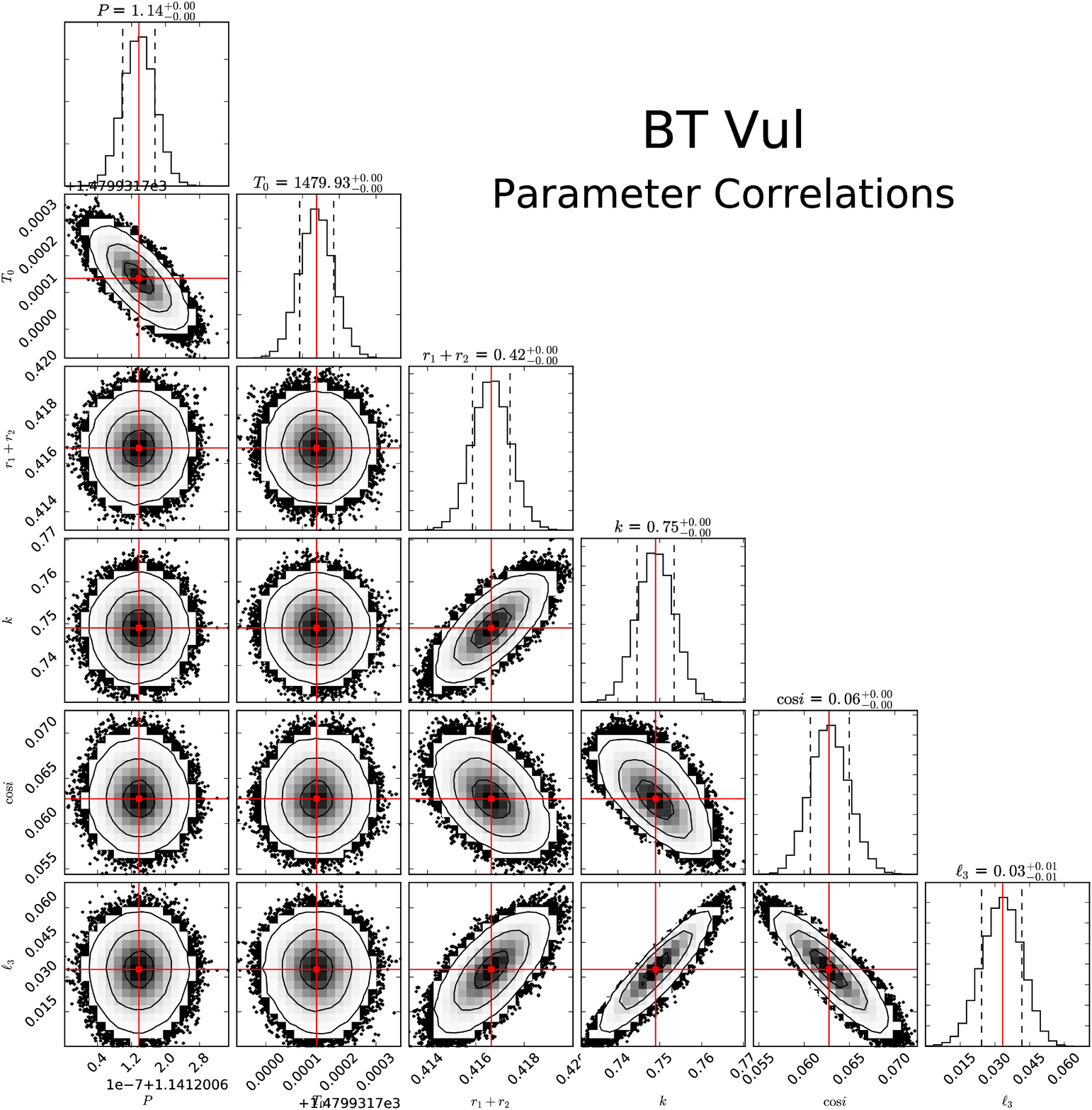}
\figcaption{``Corner plot'' \citep{Foreman-Mackey:2016}\footnote{\url
    https://github.com/dfm/corner.py~.} from the joint MCMC lightcurve
  analysis of \btvul\ illustrating the correlations among a selection
  of the fitted parameters of our solution.  Contour levels correspond
  to 1, 2, and 3$\sigma$, and the histograms on the diagonal represent
  the posterior distribution for each parameter, with the mode and
  internal 68.3\% confidence levels indicated. More realistic errors
  are discussed in the text.\label{fig:corner}}
\end{figure}

There is good agreement between the URSA and NFO results in
Table~\ref{tab:mcmc}.  The third-light parameter is seen to be
consistent with zero in both cases.  Our final solution combines the
URSA and NFO photometry, and is presented in the last column of
Table~\ref{tab:mcmc}. While most parameters are largely
  uncorrelated, a few show stronger degeneracies ($P$, $T_0$,
  $r_1+r_2$, $k$, $\cos i$, and $\ell_3$), with correlation
  coefficients among some of them being in the range of about
  0.7--0.9, in absolute value. A graphical illustration of the
  correlations among those parameters is shown in
  Figure~\ref{fig:corner}.

\begin{figure}
\epsscale{1.15}

\plotone{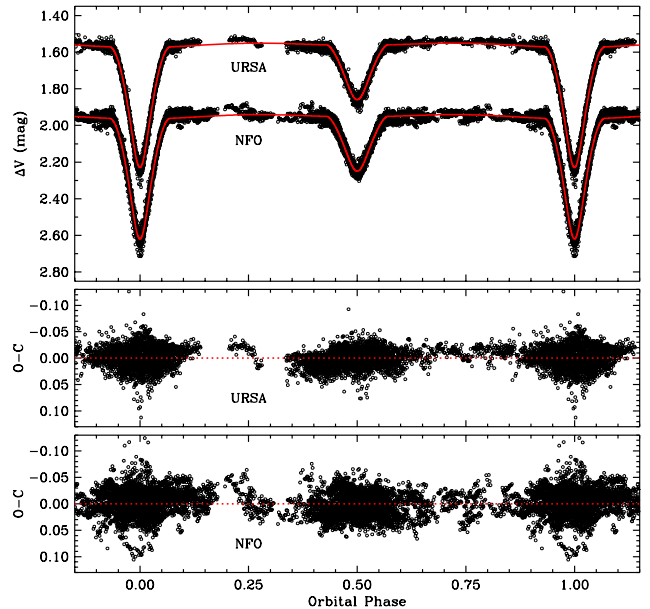}

\figcaption{URSA and NFO observations of \btvul\ along with our
  adopted lightcurve model. The NFO photometry has been displaced
  vertically for clarity. Residuals in magnitude units are displayed
  at the bottom on an expanded scale.\label{fig:lightcurves}}

\end{figure}

The oblateness of the stars, as defined by
\cite{Binnendijk:1960}, is calculated to be 0.016 for the primary and
0.011 for the secondary, which are well below the safe limit for this
binary model \citep[0.04; see, e.g.,][]{Popper:1981}.  Both eclipses
are partial. Figure~\ref{fig:lightcurves} shows the model and
photometric observations and residuals.

\setlength{\tabcolsep}{6pt}  
\begin{deluxetable*}{lcccc}
\tablewidth{0pc}
\tablecaption{Lightcurve Solutions for \btvul\ \label{tab:mcmc}}
\tablehead{
\colhead{~~~~~~~~~~~Parameter~~~~~~~~~~~} &
\colhead{URSA} &
\colhead{NFO} &
\colhead{Combined} &
\colhead{Prior}
}
\startdata
$P$ (day)\dotfill               &  1.14120067 (11)  &  1.14120088 (20)  &  1.14120073 (9)    & [1, 2] \\
$T_0$ (HJD$-$2,400,000)\dotfill &  53479.93184 (15) &  53479.93178 (29) &  53479.93184 (13)  & [53479, 53480] \\
$J$\dotfill                     &  0.532 (26)       &  0.542 (26)       &  0.537 (30)        & [0.05, 1.00] \\
$r_1+r_2$\dotfill               &  0.4149 (44)      &  0.4153 (43)      &  0.4166 (43)       & [0.05, 0.80] \\
$k \equiv r_2/r_1$\dotfill      &  0.740 (30)       &  0.733 (41)       &  0.749 (24)        & [0.1, 1.0] \\
$\cos i$\dotfill                &  0.067 (12)       &  0.066 (16)       &  0.063  (12)       & [0, 1] \\
$m_0$ URSA (mag)\dotfill        &  1.5656 (30)      &    \nodata        &  1.5650 (36)       & [1, 2] \\
$m_0$ NFO (mag)\dotfill         &     \nodata       &  1.5541 (26)      &  1.5549 (34)       & [1, 2] \\
$u_1$\dotfill                   &     0.548         &  0.548            &  0.548             & [0, 1] \\
$u_2$\dotfill                   &     0.609         &  0.609            &  0.609             & [0, 1] \\
$\ell_3$\dotfill                &  0.013 (18)       &  0.005 (16)       &  0.033 (17)        & [0.0, 0.5] \\
$f$ URSA\dotfill                &  1.077 (12)       &   \nodata         &  1.078 (15)        & [-5, 2]* \\
$f$ NFO\dotfill                 &     \nodata       &  1.455 (14)       &  1.456 (15)        & [-5, 2]* \\
$N_{\rm obs}$\dotfill           &     5451          &     6018          &  5451 + 6018       & \nodata \\ [0.5ex]
\noalign{\hrule} \\ [-1ex]
\multicolumn{5}{c}{Derived quantities} \\ [1ex]
\noalign{\hrule} \\ [-1.5ex]
$r_1$\dotfill                   &  0.2384 (32)      &  0.2396 (43)      &  0.2382 (28)       & \\
$r_2$\dotfill                   &  0.1764 (54)      &  0.1756 (67)      &  0.1784 (44)       & \\
$i$ (degree)\dotfill            &  86.13 (71)       &  86.24 (91)       &  86.40 (70)        & \\
$\ell_2/\ell_1 (V)$\dotfill     &  0.294 (22)       &  0.294 (31)       &  0.305 (17)        & 
\enddata

\tablecomments{Uncertainties are indicated in parentheses in units of
  the last significant place. They are the quadrature sum of the
    formal 68.3\% confidence intervals calculated from the posterior
    distributions, with a contribution from systematic errors based on
    the residual permutation exercise described in the text. Priors
  indicated in square brackets are uniform over the specified ranges,
  except those for $f$ (marked with asterisks), which are
  log-uniform. The limb-darkening coefficients $u_1$ and $u_2$ were
  held fixed.}

\end{deluxetable*}
\setlength{\tabcolsep}{6pt}

Because of the presence of time-correlated (``red'') noise in the NFO
photometry, and to a lesser extent also in the URSA data, in principle
there could be subtle biases in the parameters of both solutions that
would propagate through to the absolute masses, radii, and other
properties, although the good agreement found above between the two
completely independent data sets makes this appear unlikely.  A
consistency test based on the flux ratios supporting this conclusion
is presented in the next section.

An additional consequence of time-correlated noise is that the
parameter uncertainties may be underestimated, because the MCMC
procedure assumes noise is purely Gaussian. To address this concern,
we carried out a residual permutation test in which we shifted the
residuals from the URSA and NFO solutions separately by a random
number of time indices, and added them back into the model curve at
each time of observation, wrapping around at the ends of the data
sets. This preserves the time-correlated nature of the errors. These
artificial data sets were then subjected to a new MCMC solution,
simultaneously perturbing the quantities that we held fixed in the
original runs (the spectroscopic mass ratio $q$, the limb darkening
coefficients $u_1$ and $u_2$, the gravity darkening coefficients $y_1$
and $y_2$, and the albedos $A_1$ and $A_2$).  The perturbed mass
ratios were generated by adding Gaussian noise to the
spectroscopically determined value, with a standard deviation equal to
its measurement error. The other six quantities were similarly
perturbed from their adopted values with a standard deviation of 0.1.
We repeated this 50 times for each of the three solutions (URSA, NFO,
and combined), and adopted the standard deviation of the resulting
distribution for each fitted parameter as a more realistic measure of
the uncertainty. Finally, we added these uncertainties in quadrature
with the internal errors from the MCMC procedure, resulting in the
final uncertainties reported in Table~\ref{tab:mcmc}. For most
parameters the additional error contribution from red noise is
typically 2--10 times larger than the internal errors, and
occasionally even larger.

\section{Absolute dimensions}
\label{sec:dimensions}

The derived absolute masses, radii, and other properties of
\btvul\ are listed in Table~\ref{tab:dimensions}. The masses have
relative errors of about 0.6\% for both stars, and the radii are good
to 1.2\% for the primary and 2.5\% for the secondary, placing the
system among the eclipsing binaries with the best determined
properties \citep[see, e.g.,][]{Torresetal:2010}.

\setlength{\tabcolsep}{3pt}  
\begin{deluxetable}{lcc}
\tablewidth{0pc}
\tablecaption{Physical Properties of \btvul \label{tab:dimensions}}
\tablehead{ \colhead{~~~~~~~~~~Parameter~~~~~~~~~~} & \colhead{Primary} & \colhead{Secondary} }
\startdata
 $M$ ($\mathcal{M}_{\sun}^{\rm N}$)\dotfill &  $1.5439 \pm 0.0098$      &  $1.2196 \pm 0.0080$    \\
 $R$ ($\mathcal{R}_{\sun}^{\rm N}$)\dotfill &  $1.536 \pm 0.018$        &  $1.151 \pm 0.029$      \\
 $\log g$ (dex)\dotfill                     &  $4.254 \pm 0.011$        &  $4.403 \pm 0.022$      \\
 $q \equiv M_2/M_1$\dotfill                 &          \multicolumn{2}{c}{$0.7899 \pm 0.0030$} \\
 $a$ ($\mathcal{R}_{\sun}^{\rm N}$)\dotfill &          \multicolumn{2}{c}{$6.450 \pm 0.013$}   \\
 $T_{\rm eff}$ (K)\dotfill                  &  $7270 \pm 150$\phn       &  $6260 \pm 180$\phn         \\
 $L$ ($L_{\sun}$)\dotfill                   &  $5.91 \pm 0.51$          &  $1.82 \pm 0.23$        \\
 $M_{\rm bol}$ (mag)\dotfill                &  $2.804 \pm 0.093$        &  $4.08 \pm 0.14$         \\
 $BC_V$ (mag)\dotfill                       &  $-0.035 \pm 0.100$\phs   &  $-0.014 \pm 0.100$\phs     \\
 $M_V$ (mag)\dotfill                        &  $2.77 \pm 0.14$          &  $4.09 \pm 0.18$        \\
 $v_{\rm sync} \sin i$ (\kms)\tablenotemark{a}\dotfill  &  $68.0 \pm 0.8$\phn  &  $50.9 \pm 1.3$\phn   \\
 $v \sin i$ (\kms)\tablenotemark{b}\dotfill &  $70 \pm 3$\phn           &  $56 \pm 3$\phn             \\
 $E(B-V)$ (mag)\dotfill                     &          \multicolumn{2}{c}{0.148~$\pm$~0.033}   \\
 $A_V$ (mag)\dotfill                        &          \multicolumn{2}{c}{0.46~$\pm$~0.10}   \\
 Dist.\ modulus (mag)\dotfill               &          \multicolumn{2}{c}{$8.58 \pm 0.16$}     \\
 Distance (pc)\dotfill                      &          \multicolumn{2}{c}{$520 \pm 40$\phn}        \\
 $\pi$ (mas)\dotfill                        &          \multicolumn{2}{c}{$1.92 \pm 0.14$}     \\
 $\pi_{Gaia/{\rm DR2}}$ (mas)\tablenotemark{c}\dotfill  &  \multicolumn{2}{c}{$1.872 \pm 0.049$} 
\enddata
\tablecomments{The masses, radii, and semimajor axis $a$ are expressed
  in units of the nominal solar mass and radius
  ($\mathcal{M}_{\sun}^{\rm N}$, $\mathcal{R}_{\sun}^{\rm N}$) as
  recommended by 2015 IAU Resolution B3 \citep[see][]{Prsa:2016}, and
  the adopted solar temperature is 5772~K (2015 IAU Resolution
  B2). Bolometric corrections are from the work of \cite{Flower:1996},
  with conservative uncertainties of 0.1~mag, and the bolometric
  magnitude adopted for the Sun appropriate for this $BC_V$ scale is
  $M_{\rm bol}^{\sun} = 4.732$ \citep[see][]{Torres:2010}. See text
  for the source of the reddening. For the apparent visual magnitude
  of \btvul\ out of eclipse we used $V = 11.49 \pm 0.01$
  \citep{Henden:2015}.}
\tablenotetext{a}{Synchronous projected rotational velocity assuming
  spin-orbit alignment.}
\tablenotetext{b}{Spectroscopically measured projected rotational
  velocities.}
\tablenotetext{c}{A global parallax zero-point correction of
  $+0.029$~mas has been added to the original {\it Gaia}/DR2 parallax
    \citep{Lindegren:2018a}, and 0.021~mas added in quadrature to the
    internal error \citep[see][]{Lindegren:2018b}.}
\end{deluxetable}
\setlength{\tabcolsep}{6pt}  

Effective temperatures are more challenging to determine accurately
than the masses or radii. As indicated in
Section~\ref{sec:spectroscopy}, we are only able to derive a
spectroscopic value for the primary star, and the secondary $T_{\rm
  eff}$ scales directly with it.  Degeneracies with surface gravity
and metallicity in the way we determine the primary $T_{\rm eff}$ make
it sensitive to those quantities, such that increasing $\log g$ or
[Fe/H] by 0.5~dex results in a temperature between 200 and 300~K
hotter. To our knowledge there is no spectroscopic determination of
the metallicity of \btvul, which would be challenging because of the
significant line broadening in both stars. While the radial-velocity
determinations from the CfA spectra do not typically require knowledge
of $\log g$ or [Fe/H] to much better than 0.5~dex, for the highest
accuracy in the absolute dimensions we have chosen to interpolate the
primary $T_{\rm eff}$ to the final values of $\log g$
(Table~\ref{tab:dimensions}) and [Fe/H] (see next section).  We
obtained $7270 \pm 150$~K, which we report in the table.  The
temperature difference between the primary and secondary may be
derived from the disk-averaged surface brightness ratio (a function of
$J$ and the limb-darkening coefficients) through the use of the visual
absolute flux scale of \cite{Popper:1980}. The result, $\Delta T_{\rm
  eff} = 1010 \pm 110$~K, combined with the primary temperature, leads
to an estimate for the secondary of $T_{\rm eff} = 6260 \pm 180$~K,
which is listed also in Table~\ref{tab:dimensions}. These temperatures
correspond to spectral types of approximately F0 and F7, according to
the tabulation by \cite{Pecaut:2013}.

While the measured projected rotational velocity for the primary of
\btvul\ agrees with the value expected if the star's spin were
synchronized and aligned with the orbital motion, the secondary's $v
\sin i$ value appears slightly higher than predicted, although the
difference is small and may not be significant.  Tidal theory predicts
both stars in a binary with a period as short as this should
synchronize on a timescale much less than 1~Myr \citep[see,
  e.g.][]{Hilditch:2001}.

Interstellar reddening toward \btvul\ was estimated from the 3-D
extinction map of
\cite{Green:2019}\footnote{\url{http://argonaut.skymaps.info/}}, and
is $E(B-V) = 0.148 \pm 0.033$~mag.  With the corresponding visual
extinction, $A_V = 0.46 \pm 0.10$~mag (assuming $R_V = 3.1$), along
with the apparent visual magnitude of the system out of eclipse
\citep[$V = 11.49 \pm 0.01$;][]{Henden:2015} and bolometric
corrections from \cite{Flower:1996}, we infer a distance of $520 \pm
40$~pc. This is in very good agreement with the distance of $534 \pm
14$~pc inferred from the trigonometric parallax listed in the {\it
  Gaia}/DR2 catalog, after a small correction to the published value
and to its uncertainty as explained in Table~\ref{tab:dimensions}.

\begin{figure}
\epsscale{1.15}
\plotone{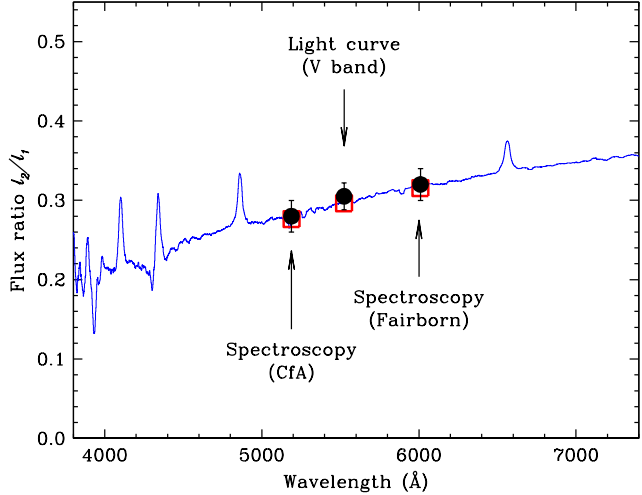}

\figcaption{Flux ratio $\ell_2/\ell_1$ as a function of wavelength,
  calculated using solar-metallicity synthetic spectra by
  \cite{Husser:2013} for temperatures of 7300~K (interpolated) and
  6300~K near those adopted for the primary and secondary, and $\log
  g$ values of 4.0 and 4.5, respectively. The normalization was
  performed with a radius ratio $k = 0.749$ from our lightcurve
  solution (Table~\ref{tab:mcmc}). Squares indicate the predicted
  values integrated over the corresponding bandpasses. The ratios
  measured spectroscopically and from the light curves (points with
  error bars) are shown to agree well with the expected
  values.\label{fig:fluxratio}}

\end{figure}

Our spectroscopic measurements of the flux ratio between the
components present an opportunity for a check on the internal
consistency of our analysis. Neither of those two empirical values of
$\ell_2/\ell_1$ is at the same wavelength as the photometry, so a
direct comparison with the $V$-band ratio from the lightcurve analysis
is not possible. In its place, we have used synthetic spectra based on
PHOENIX models from the library of \cite{Husser:2013} for temperatures
near our adopted values for the components, in order to predict the
flux ratio as a function of wavelength. This is seen in
Figure~\ref{fig:fluxratio}, where the normalization of the ratio of
the stellar fluxes was carried out using our photometric radius ratio
$k = 0.749$. Both spectroscopic flux ratios, as well as the one
derived at $V$ from the light curve, agree very well with the
prediction, within their uncertainties, supporting the accuracy of our
determinations.

\section{Comparison with stellar evolution models}
\label{sec:models}

\begin{figure}
\epsscale{1.15} \plotone{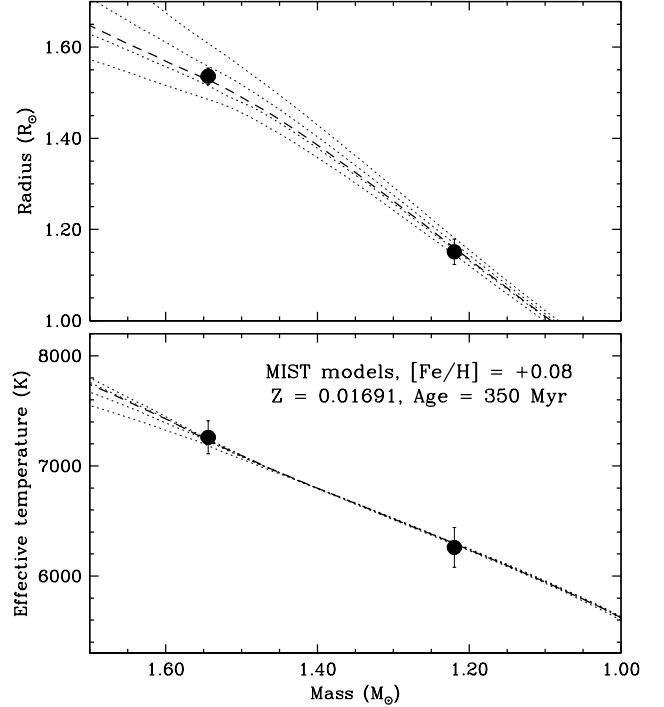}

\figcaption{Mass-radius and mass-temperature diagrams, comparing our
  determinations for \btvul\ against model isochrones from the MIST
  series \citep{Choi:2016} for a metallicity of ${\rm [Fe/H]} = +0.08$
  that best fits the measurements. Isochrones between 100 and 700~Myr
  in steps of 200~Myr are indicated with dotted lines, with the best
  fit represented with a heavy dashed line for an age of
  350~Myr. \label{fig:massradius}}

\end{figure}

Our mass, radius, and temperature determinations for the components of
\btvul\ are displayed in the mass-radius and mass-temperature diagrams
of Figure~\ref{fig:massradius}, where they are compared against
stellar evolution models from the MIST series \citep{Choi:2016}. The
dotted lines represent model isochrones for ages between 100 and
700~Myr, in steps of 200~Myr. The best-fit age of 350~Myr is
represented with a heavy dashed line. The model metallicity adopted
for this comparison (assuming no $\alpha$-element enhancement) is
${\rm [Fe/H]} = +0.08$, which is the abundance that provides the best
match to the effective temperatures. Note that the empirical $T_{\rm
  eff}$ values are tied to our spectroscopic determination for the
primary from the CfA observations, which, as explained in
Section~\ref{sec:spectroscopy}, is in turn dependent on the chemical
composition. We therefore proceeded by iterations, changing the model
abundance and then adjusting the spectroscopic temperatures
accordingly until reaching agreement between the predicted and
observed temperatures at the same composition. There is very good
consistency between theory and our $M$, $R$, and $T_{\rm eff}$
determinations.

\begin{figure}
\epsscale{1.15}
\plotone{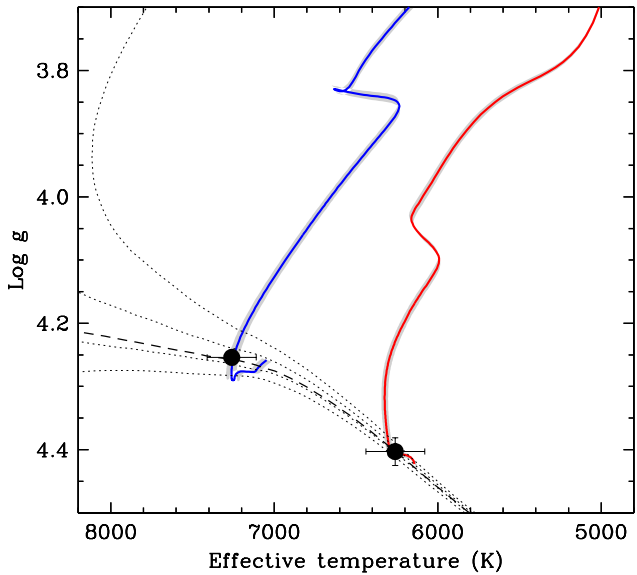}

\figcaption{Properties for \btvul\ shown against evolutionary tracks
  from \citep{Choi:2016}, computed for the exact masses we measure and
  ${\rm [Fe/H]} = +0.08$. The shaded areas around each model
  correspond to the uncertainty in the location of the track that
  comes from the mass errors. Dotted and dashed lines correspond to
  the same isochrones shown in Figure~\ref{fig:massradius}. The
  components of \btvul\ are seen to be near the
  ZAMS. \label{fig:tracks}}

\end{figure}

The models point to a fairly young system. The evolutionary state of
\btvul\ is illustrated more clearly in the Kiel diagram of
Figure~\ref{fig:tracks}, in which evolutionary tracks are shown for
the exact masses we measure.  Both stars are seen to be very near the
zero-age main sequence (ZAMS). The dotted and dashed lines represent
the same isochrones from Figure~\ref{fig:massradius}.

\section{Final remarks}
\label{sec:remarks}

Our photometric and spectroscopic monitoring of the neglected F-type
detached eclipsing binary \btvul\ have allowed us to accurately
characterize the components, and to measure their masses to a
precision of about 0.6\% and their radii to 1.2 and 2.5\% for the
primary and secondary. The comparison with stellar evolution models
indicates a system age of about 350~Myr, both stars being very near
the ZAMS. Our distance estimate (520~pc) and center-of-mass velocity,
along with the {\it Gaia\/} proper motion and position, imply space
velocity components of [$U$, $V$, $W$] = [$-1.2$, $-23.4$,
  $-9.5$]~\kms, which are typical of the thin disk in the Milky Way
and are consistent with the young age and near-solar composition we
infer.

In an astrometric survey of solar-type spectroscopic binaries for
additional companions, \cite{Tokovinin:2006} have estimated that the
vast majority (as many as 96\%) of short-period binaries with periods
under 3~days are in fact triple or higher-order systems. This is
interpreted as evidence for the importance of dynamical processes at
play early-on in the evolution of multiple systems, such as Kozai
cycles with tidal friction \citep[e.g.,][]{Eggleton:2001,
  Fabrycky:2007}. These mechanisms gradually tighten the inner binary
through angular momentum exchange with the third star.

Given its orbital period of 1.14~days, it would not be surprising if
\btvul\ turned out to be a triple system as well, although there is
currently no compelling evidence for this in the observations at hand.
We detect no long-term trend in the velocity residuals from our
spectroscopic orbital solution, and although our final lightcurve
solution does return a non-zero value for the third-light parameter
($\ell_3 = 0.033 \pm 0.017$) that could be caused by a tertiary star,
the measurement is only marginally significant. Additionally, we have
examined our CfA spectra with TRICOR \citep{Zucker:1995}, an extension
of TODCOR to three dimensions, and we see no indication of a third set
of lines in our spectra.

The {\it Gaia}/DR2 catalog reports a small level of excess astrometric
noise in their 5-parameter solution for \btvul\ of 0.090~mas, with a
dimensionless statistical significance of $D = 3.83$
\citep[see][]{Gaia:2018}. This is somewhat above the adopted threshold
for the catalog that would indicate real unmodeled effects ($D = 2$).
While this extra noise could in principle be due to motion in a triple
system, it is currently not possible to rule out instrumental effects.
High-resolution imaging with adaptive optics in the near infrared,
where the contrast with a presumably late-type tertiary star would be
more favorable, could shed more light on this issue.

\acknowledgments

The spectroscopic observations of \btvul\ at the CfA were gathered
with the expert assistance of P.\ Berlind, M.\ Calkins, G.\ Esquerdo,
D.\ Latham, and R.\ Stefanik. We also thank J.\ Mink for maintaining
the CfA echelle database, as well as Bill Neely, who operated and
maintained the NFO WebScope for the Consortium and who handled
preliminary processing of the images and their distribution. The
  anonymous referee is thanked for helpful comments on the original
  manuscript.  G.T.\ acknowledges partial support from the National
Science Foundation (NSF) through grant AST-1509375. Astronomy at
Tennessee State University is supported by the state of Tennessee
through its Centers of Excellence Program.  This research has made use
of the SIMBAD and VizieR databases, operated at the CDS, Strasbourg,
France, and of NASA's Astrophysics Data System Abstract Service. The
work has also made use of data from the European Space Agency (ESA)
mission {\it Gaia\/} (\url{https://www.cosmos.esa.int/gaia}),
processed by the {\it Gaia\/} Data Processing and Analysis Consortium
(DPAC,
\url{https://www.cosmos.esa.int/web/gaia/dpac/consortium}). Funding
for the DPAC has been provided by national institutions, in particular
the institutions participating in the {\it Gaia} Multilateral
Agreement. The computational resources used for this research include
the Smithsonian Institution's ``Hydra'' High Performance Cluster.

\end{document}